\documentclass[twocolumn]{aastex63}

\usepackage{graphicx}
\usepackage{txfonts}
\usepackage[utf8]{inputenc}
\usepackage{natbib}
\usepackage{float}
\usepackage{mathrsfs}
\usepackage{hyperref}

\received{October 12, 2020}
\revised{October 22, 2020}
\accepted{October 23, 2020}
\submitjournal{The Astrophysical Journal}

\shorttitle{Coronal electron densities from the 21 August 2017 total solar eclipse}
\shortauthors{Bemporad}

\begin{document}

\title{Coronal Electron Densities derived with Images acquired\\
during the 21 August 2017 Total Solar Eclipse}

\author[0000-0001-5796-5653]{A. Bemporad}
\affil{Istituto Nazionale di Astrofisica (INAF), Osservatorio Astrofisico di Torino, \\ via Osservatorio 20, 10025 Pino Torinese, Torino, Italy} 
\email{alessandro.bemporad@inaf.it}

\begin{abstract}
The total solar eclipse of August 21st, 2017 was observed with a Digital Single Lens Reflex (DSLR) camera equipped with a linear polarizing filter. A method was developed to combine images acquired with 15 different exposure times (from 1/4000 sec to 4 sec), identifying in each pixel the best interval of detector linearity. The resulting mosaic image of the solar corona extends up to more than 5 solar radii, with a projected pixel size by 3.7 arcsec/pixel, and an effective image resolution by 10.2 arcsecs, as determined with visible $\alpha-$Leo and $\nu-$Leo stars. Image analysis shows that in the inner corona the intensity gradients are so steep, that nearby pixels shows a relative intensity difference by up to $\sim 10 \%$; this implies that careful must be taken when analyzing single exposures acquired with polarization cameras.

Images acquired with two different orientations of the polarizer have been analyzed to derive the degree of linear polarization, and the polarized brightness $pB$ in the solar corona. After inter-calibration with $pB$ measurements by the K-Cor instrument on Mauna Loa Solar Observatory (MLSO), data analysis provided the 2D coronal electron density distribution from 1.1 up to $\sim 3$ solar radii. The absolute radiometric calibration was also performed, with the full sun image, and with magnitudes of visible stars. The resulting absolute calibrations show a disagreement by a factor $\sim 2$ with respect to MLSO; interestingly, this is the same disagreement recently found with eclipse predictions provided by MHD numerical simulations. 
\end{abstract}


\section{Introduction}
Despite the actual availability of coronagraphic data acquired by many ground-based and space-based instruments, the occurrence of total solar eclipses still offers today a unique opportunity to observe the full corona from almost the edge of the solar disk up to many solar radii, allowing to test new instrumentation \citep[e.g.][]{samra2018, madsen2019}, new ideas \citep{reginald2019}, and complementing other observations \citep{pasachoff2017}. Moreover, these fascinating events offers at the same time the possibility to involve the general public in astrophysics in general, and solar physics in particular.

Over the last $\sim 15$ years, the actual availability of standard Digital Single Lens Reflex (DSLR) cameras, coupled with personal computers and freeware astro-imaging tools, allowed an increasing number of people (scientists, amateur astronomers, teachers) not only to acquire high-quality astronomical observations, but also to perform real scientific research. A nice example is given by the increasing number of papers analyzing these images to perform for instance stellar photometry \citep{hoot2007, pieri2012, kloppenborg2012, zhang2016, axelsen2017}. Thanks to the linearity of more recent CCDs mounted on DSLR cameras, it has been shown that these instruments can be used to characterize variable stars and novae \citep{fiacconi2009, collins2009, loughney2010, banys2014, deshmukh2015, walker2015, pyatnytskyy2019, nesci2020} even without the need for a telescope and a mount motor drive, but also to observe the transits of exo-planets \citep{littlefield2010, miller2015}, eclipsing binary stars \citep{collins2013, richards2019}, meteor spectral emissions \citep{cheng2011}, asteroid occultations \citep{hoot2012}, and even to build color-magnitude diagrams of open clusters \citep{jang2015}. Although these works are mostly addressed to amateur astronomers, teachers and educators, they also showed that, even though not explicitly designed for scientific applications, DSLR cameras can nevertheless produce high-quality data only a minimal investment of funds.

Surprisingly, the above list of published papers shows a very limited number of works doing research with DSLR cameras in solar physics, with the exception of a few recent works focusing on the measurement of plasma physical parameters in a quiescent prominence \citep{jeicic2014} based on the method by \citet{jeicic2009}, the determination of contact times of solar eclipses and planetary transits across the solar disk \citep{digiovanni2016}, the measurement of the apparent variations of the size of the Sun \citep{trillenberg2019}, and the recent observations of the total solar eclipse (TSE) of 21st August 2017 \citep{pasachoff2018, snik2020}. In particular, the latter spectacular event was observed by thousands of people as the path of totality crossed the whole US country from coast to coasts, and allowed for the first time to involve the general public on vast citizen science projects, for instance to observe cloud and temperature properties associated with the transit of the eclipse \citep{dodson2019}, to measure the ionospheric response to the variable solar illumination \citep{frissell2018}, to capture \citep[with the "Citizen CATE Experiment";][]{penn2020} a time sequence of white-light coronal observations with identical instruments over $\sim 90$ minutes of totality, or to collect \citep[with the "Eclipse Megamovie Project";][]{hudson2011} all DSLR pictures of the solar eclipse acquired by people across the US to create a movie showing the high-resolution coronal dynamics close to the limb \citep[see][for first results]{hudson2018, peticolas2019}. 

Almost $\sim 100$ research papers have been already published on the 21st August 2017 TSE, dealing with data acquired from the ground with professional instrumentation and equipment, and studying many different aspects such as the occurrence of transient and dynamic events \citep[e.g.][]{hanaoka2018, boe2020, filippov2020}, spectroscopic emissions by the E- and F-corona \citep[e.g.][]{pasachoff2018, samra2018, koutchmy2019, judge2019}, validation of MHD models \citep[][]{nandy2018, mikic2018, lamy2019}, and many other topics related not only with research on solar physics, but also on the response of the Earth's ionosphere and atmosphere \citep[e.g.][]{reinisch2018}. On the other hand, only a couple of works discussed the scientific research that can be conducted simply with DSLR cameras, to constrain the locations of fainter coronal structures \citep{pasachoff2018}, and to measure the degree of linear polarization \citep{snik2020}. 

In this work I demonstrate how images acquired during TSE with a single basic DSLR camera equipped with cost effective ND filter and linear polarizer can be analyzed to derive not only beautiful high-resolution images of the corona, but also to calibrate the polarized emission and measure the coronal electron densities. After a first description of instrumentation and observations (\S\ref{sec: observations}), I will describe how the images have been analyzed, calibrated, and combined in mosaics (\S\ref{sec: mosaic}), and discuss some interesting results from the mosaics (\S\ref{sec: RGBanalysis}). Then, I will focus on the analysis of images acquired with the linear polarizer (\S\ref{sec: POLanalysis}), and show how, after relative and absolute radiometric calibrations (\S\ref{sec: radiometric}), the coronal electron densities have been finally measured (\S\ref{sec: densities}); results are then summarized (\S\ref{sec: summary}).
\begin{figure}[t!]
\centering
\includegraphics[width=0.5\textwidth]{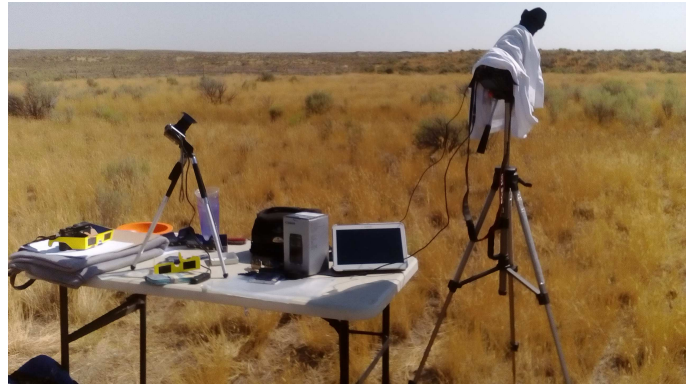}
\caption{
A picture showing the equipment employed on August 21st, 2017 to acquire the TSE observations analyzed here (see text); the picture shows the DSLR camera mounted on the tripod, connected with remote controllers, and covered by a white tissue to reduce as more as possible the overheating by solar illumination during PSE.
}
\label{Fig1}
\end{figure}

\section{Instrumentation and observations} \label{sec: observations}

A preliminary description of the observational campaign is provided in \cite{bemporad2017}. The images were acquired nearby Idaho Falls, in a location where the expected duration of totality was 2 minutes and 18 seconds; local seeing was almost perfect, with no visible clouds of any kind over the whole sky. The observations (Fig.~\ref{Fig1}) were performed with a Canon EOS 1100D DSLR Camera, equipped with a EF 75-300mm f/4-5.6 III telephoto zoom lens, and mounted on a fixed tripod (alto-azimuthal mount, no tracking); both the partial and total solar eclipse phases have been observed. In particular, the two partial solar eclipse (PSE) phases (both before and after the totality) where fully covered by using a Baader OD5.0 solar filter (mounted on a sunshade), and setting the exposure time $t_{exp}$ to 1/4000 sec, F-stop f/5.6, ISO 100 sensitivity. The camera was first hand focused at the maximum available focal length (300 mm) by looking at the edge of the Sun and also at the few small sunspots that were visible on-disk; the same focus has then been employed for both the PSE and TSE phases observations.

\begin{figure*}[t!]
\centering
\includegraphics[width=0.7\textwidth]{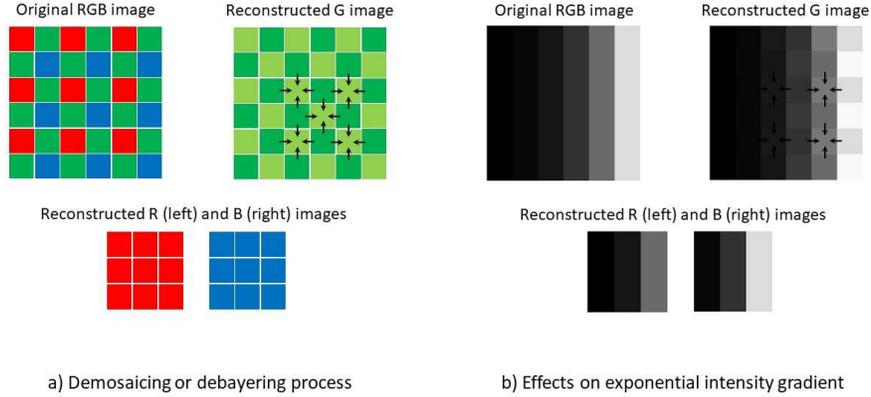}
\caption{
Panel a: schematic representation of the methods followed to reconstruct from the original RGB image (top left) acquired with RGB Bayer filter the three images in the G (top right), R and B channels (bottom right); black arrows indicate the averaging performed among nearby green pixels to reconstruct the whole G image at full resolution. Panel b: effects of the applied methods by assuming a uniform left-right exponential intensity gradient in the original image (top left), and the gradient in the resulting images (right); notice the systematic difference between the intensities of reconstructed R and B images. 
}
\label{Fig2}
\end{figure*}
With the help of a programmable LCD Digital Timer Remote Control, exposures were acquired at a time step by 68 sec from the beginning of PSE (first moon contact C1) to 2 min and 45 sec before beginning of TSE (second moon contact C2), then the time step was increased to 3 sec from C2 to the beginning of TSE. The reason for this was to measure the shape of PSE illumination curve as a function of time from C1 to C2 to test the theoretical curve for penumbra illumination level as derived for the ESA PROBA3 project \citep[see][]{bemporad2015}. Results from the analysis of these images (and those acquired from the end of TSE at moon contact C3 to the end of PSE at moon contact C4) will be described in a future publication.

At the beginning of TSE the OD5.0 filter was quickly dismounted, and a first set of TSE exposures was acquired. In particular, auto-bracketing has been performed by connecting the DSLR camera with a tablet running a freeware DSLR Controller app. A first sequence of 15 exposures (with 15 different exposure times going from $t_{exp} = 1/4000$ sec up to $t_{exp} = 4$ sec, see later on Fig.~\ref{Fig5}) was acquired, requiring a total acquisition time of about 35 sec for the whole sequence. Then, a linear polarizer filter (Hoya 58mm B58PLGB) was mounted in front of the zoom and (after verification of the orientation of the linear polarizer with respect to a reference mark) a second sequence of 15 exposures was acquired. The orientation of the linear polarizer was rotated clockwise by $\sim 90^\circ$ and a third sequence of 15 exposures was acquired again. Finally, after a further counter-clockwise rotation of the polarizer by $\sim 45^\circ$, a fourth and last sequence was acquired. Between each polarized sequence the orientation of the linear polarizer was rotated by moving a reference arrow with respect to grooves on the rotating part of the filter mount (marked before the observational campaign) and separated by the right angular distances. At the end of TSE the linear polarizer was dismounted, and the OD5.0 filter was mounted again starting the acquisition of the second PSE sequence first with a time step by 3 sec, and then with a time step by 68 sec to the end of PSE.

At the end of the observations, three sequences of 15 dark frames were acquired by covering the zoom with the cap and by employing exactly the same $t_{exp}$ used to acquire the bracketing sequence during TSE. The same was also repeated for flat field images, that were acquired by covering the lens with a uniform white fabric and pointing the camera to the sky.

\section{From image sequences to mosaics} \label{sec: mosaic}

The first step in the analysis of images acquired with a DSLR camera is the conversion from the RAW files to another format that is readable for the analysis by any programming language. In particular, in this work the images have been converted from RAW to TIFF formats with the open source program \href{https://en.wikipedia.org/wiki/Dcraw}{DCRAW} freely distributed on-line. For the rest of the analysis described here all the routines have been written in \href{https://en.wikipedia.org/wiki/IDL_(programming_language)}{IDL} language, but any other open source programming language (such as Python or others) could be used.
\subsection{Image extraction}
\begin{figure}[b!]
\centering
\includegraphics[width=0.45\textwidth]{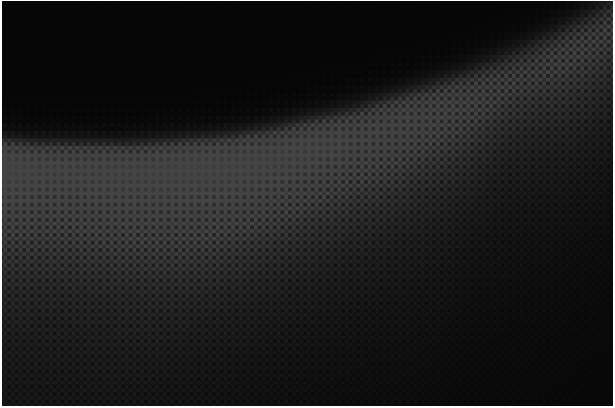}
\caption{
A zoom over the region near the edge of the occulting Moon in one of the RAW images acquired during the first TSE sequence; the image shows clearly the Bayer alternating pattern of RGB pixels.  
}
\label{Fig3}
\end{figure}

\begin{figure*}[t!]
\centering
\includegraphics[width=0.9\textwidth]{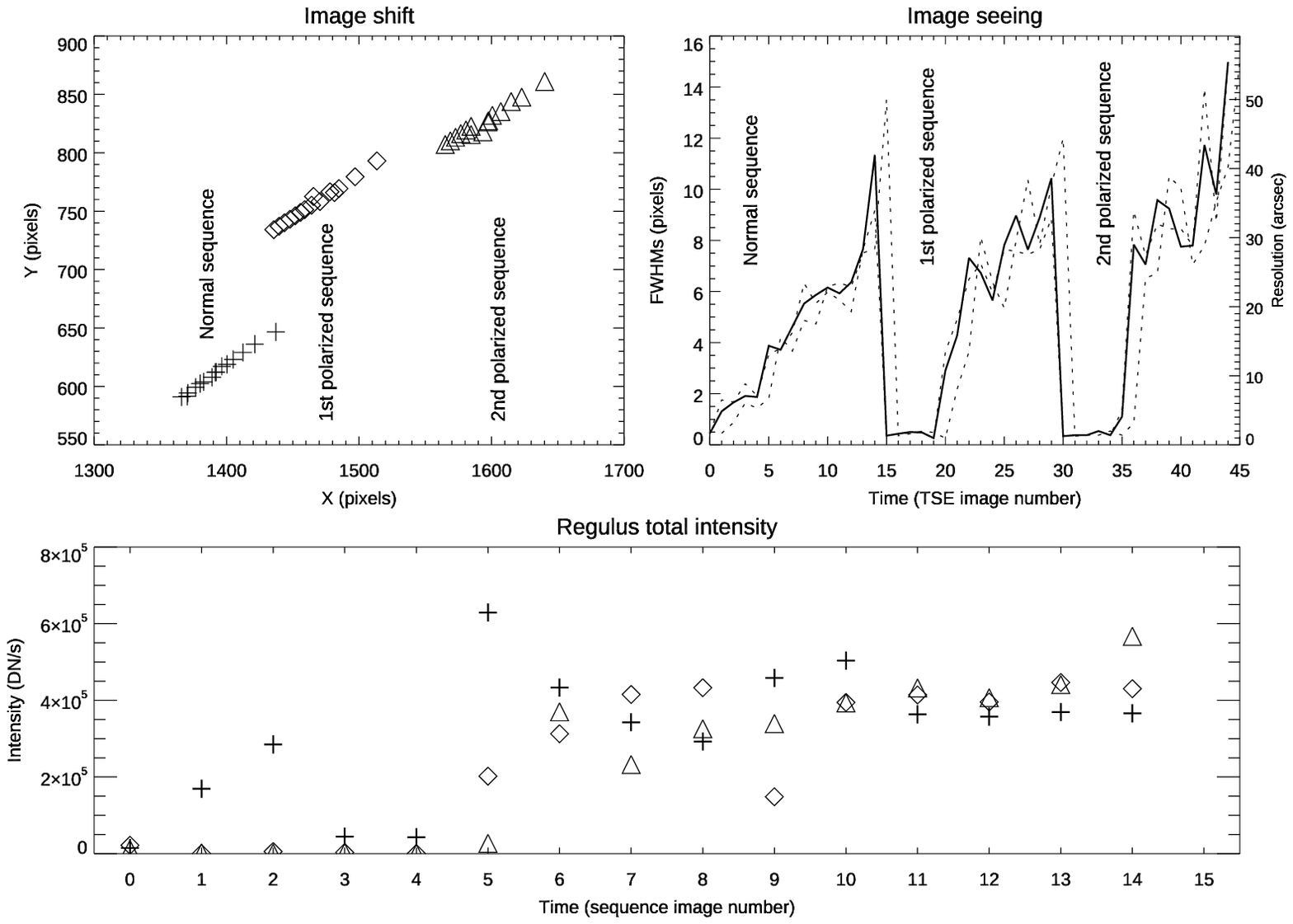}
\caption{
Top left: derived location (in pixels) of the Regulus star in each frame for the three acquired sequences (see text). Top right: corresponding FWHMs in the x and y directions (dotted lines) as derived from bi-dimensional Gaussian fitting, and the average FWHM (solid line) for the three image sequences. Bottom: corresponding total intensity [DN/s] normalized to the exposure time $t_{exp}$ for the Regulus star for each image in the three sequences (symbols are the same used in the top left plot). Fluxes from polarized sequences have been multiplied by a factor 4 to be comparable with those acquired without the polarizer.  
}
\label{Fig4}
\end{figure*}
The second step in the analysis consists in the so-called demosaicing or debayering process \citep[see e.g. ][]{pata2010}: all DSLR cameras acquire images with a digital sensor overlaid with a color filter array (CFA) which is usually a Bayer filter alternating red (R) and green (G) filters for odd rows and green (G) and blue (B) filters for even rows (Fig.~\ref{Fig2}). Because different filters (hence different pixels) are integrating over different wavelength intervals, the intensity in each pixel also depends on the RGB color filter, and for the scientific analysis it is necessary to separate the three RGB colors. Hence, each TIFF image created from the RAW file (Fig.~\ref{Fig3}) has been converted into three separate images for each one of the three RGB colors. The TIFF images have (4272 $\times$ 2848) pixels, but with a Bayer filter 1/2 of pixels have a G filter, while 1/4 of pixels have a B or a R filter (Fig.~\ref{Fig2}). Hence, G images have been constructed with the same number of (4272 $\times$ 2848) pixels by simple interpolation, by replacing values in each R or B pixel with the average between the 4 nearby G pixels (black arrows in Fig.~\ref{Fig2}, panel a). On the other hand, R and B images have been constructed simply by reading the (2136 $\times$ 1424) pixels with the R and B filters. This procedure has been applied not only to images acquired during the PSE and TSE observations, but also to dark and flat field images. The effects of these reconstruction methods in the coronal intensity gradients reconstructed in the three channels (Fig.~\ref{Fig2}, panel b) will be discussed later.

\subsection{Image correction}

Starting from the acquired dark and flat field images, master images have been created, and all images have been corrected both for the dark currents and the flat field. The 2D distribution of flat field intensities also provides correction for the vignetting of the employed optical system, that was not very important: the normalized flat field images show center-to-corner relative intensity decreases by less than $\simeq 25\%$. Because the used tripod was not motorized to follow the motion of the sky during the eclipse, before combination of different exposures image co-alignement is required. This is a very important step, and the edge of the occulting disk of the moon cannot be used as a reference for co-alignement, because during the TSE the Moon is moving with respect to the Sun, and the images need to be co-aligned with respect to the center of the Sun and not to the center of the Moon.

The easiest way to co-align different images is to use visible stars. In the G images two brighter stars were clearly visible and were identified (by using the free open source program \href{http://stellarium.org/}{Stellarium}): Regulus or $\alpha-$Leonis (apparent visual magnitude +1.35), and $\nu-$Leonis (apparent visual magnitude +5.15), located respectively bottom left and top right with respect to the Sun. The images have been then co-aligned by using the positions of the brightest star Regulus determined in each frame. In the first sequence of 15 images acquired without the polarizer the star intensity is sufficiently high to identify its position in all the frames acquired with $t_{exp} > 1/60$ sec. The positions of the star in the first frames were derived by back-extrapolating in time (with linear fitting) the position of the star derived with longer exposures. Results (top left panel in Fig.~\ref{Fig4}) show that during the first sequence the star displaced by about 90.3 pixels (in $x$ and $y$ directions). This displacement corresponds to 334.3 arcsec, considering a pixel projected size by 3.7 arcsec/pix. Because the acquisition of the whole sequence of 15 exposures required about 35 sec, this corresponds to an average motion of the sky by 9.5 arcsec/sec, corresponding to $\sim 2.5$ pixels/sec. 

The above pixel projected size was determined from images acquired during the PSE: the Sun and Moon disks were fitted with circles, providing values of $R_{sun} = (257.7 \pm 0.2)$ pix and $R_{moon} = (267.0 \pm 0.2)$ pix for the projected radii of the Sun and the Moon. Because on 21st August 2017 the Sun was at a distance of 1.012 AU from the Earth, the projected radius covered 948.7 arcsec, hence the pixel size was 948.7/257.7 = 3.7 arcsec/pixel, corresponding to a two-pixel resolution by 7.4 arcsec. The acquisition of first 5 frames required only $\sim 5$ secs, hence the expected motion of the star during the first 5 exposures was on the order of $\sim 13$ pixels. The same method was applied to determine the position of the star Regulus during the 1st and 2nd sequences acquired with the polarizer. Each sequence has been treated separately, because the application of the polarizing filter on the camera (between the first and the second sequence) and the rotation of the filter (between the second and the third sequence) slightly changed the pointing of the camera, leading to the discontinuities visible in Fig.~\ref{Fig4} (top left panel).

The centroid position of the star in each frame was determined with bi-dimensional Gaussian fitting, and this also provided an estimate of the effective image resolution (given mainly by the combination of local seeing and the PSF of the optical system). Resulting values of FWHMs for each one of the three sequences are shown in Fig.~\ref{Fig4} (top right panel). Because of the motion of the sky during the acquisition times, the measured FWHMs increase as $t_{exp}$ become longer in each sequence; hence a reference value is provided by the FWHM measured in the image acquired with the longer $t_{exp}$ (to have a better signal-to-noise ratio), but still smaller than the time required to the star to move significantly in the image, for instance by more than 1/4 of pixel ($\sim 0.1$ sec). This corresponds to the image acquired with $t_{exp} = 1/16$ sec: from this image the measured FWHM of Regulus turns out to be 5.52 pixels (hence HWHM by 2.76 pixels), corresponding to an effective image resolution (seeing plus PSF) by 10.2 arcsecs. Images acquired with longer exposure times were affected by blurring due to the motion of the sky, with higher effective resolutions shown in Fig.~\ref{Fig4} (top right panel).

The bi-dimensional Gaussian fitting of Regulus also provides an estimate of the total intensity $I_{G,\alpha}$ in the G channel normalized for the exposure time (DN/sec). In particular, the intensities measured for all exposures acquired in the three sequences are shown in Fig.~\ref{Fig4} (bottom panel), where the values measured with the polarizer (diamond and triangle symbols) have been multiplied by a factor 4 to be comparable with values measured without the polarizer (plus symbols). The resulting intensities measured with the G images are almost constant for different exposure times, and the intensity is on average $I_{G,\alpha} = (3.64 \pm 0.05) \times 10^5$ DN/sec. This means that (at least for Regulus which is a relatively weak source if compared with the much brighter inner solar corona) the linearity of the detector response is good, a characteristic which is very important to combine all the exposures, as explained below.

\begin{figure}[b]
\centering
\vspace{-0.9cm}
\includegraphics[width=0.49\textwidth]{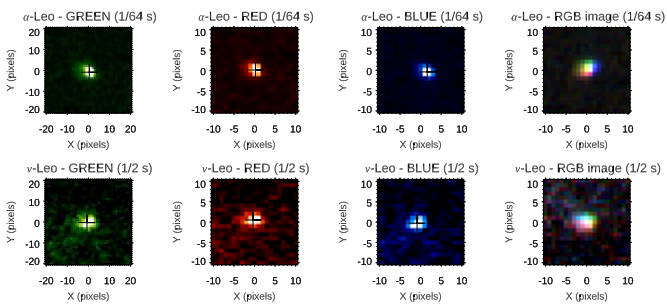}
\vspace{-1.3cm}
\caption{
A zoom over the intensity distributions of the $\alpha-$Leonis (top row) and $\nu-$Leonis (bottom row) stars as obtained from the G (left), R (middle left), and B (middle right) channels, and the resulting RGB combined images (right panels) showing chromatic aberrations of the optical system. Plus symbols show the centroid location of intensity distribution for each color.
}
\label{Fig65}
\end{figure}
\begin{figure*}[t!]
\centering
\includegraphics[width=\textwidth]{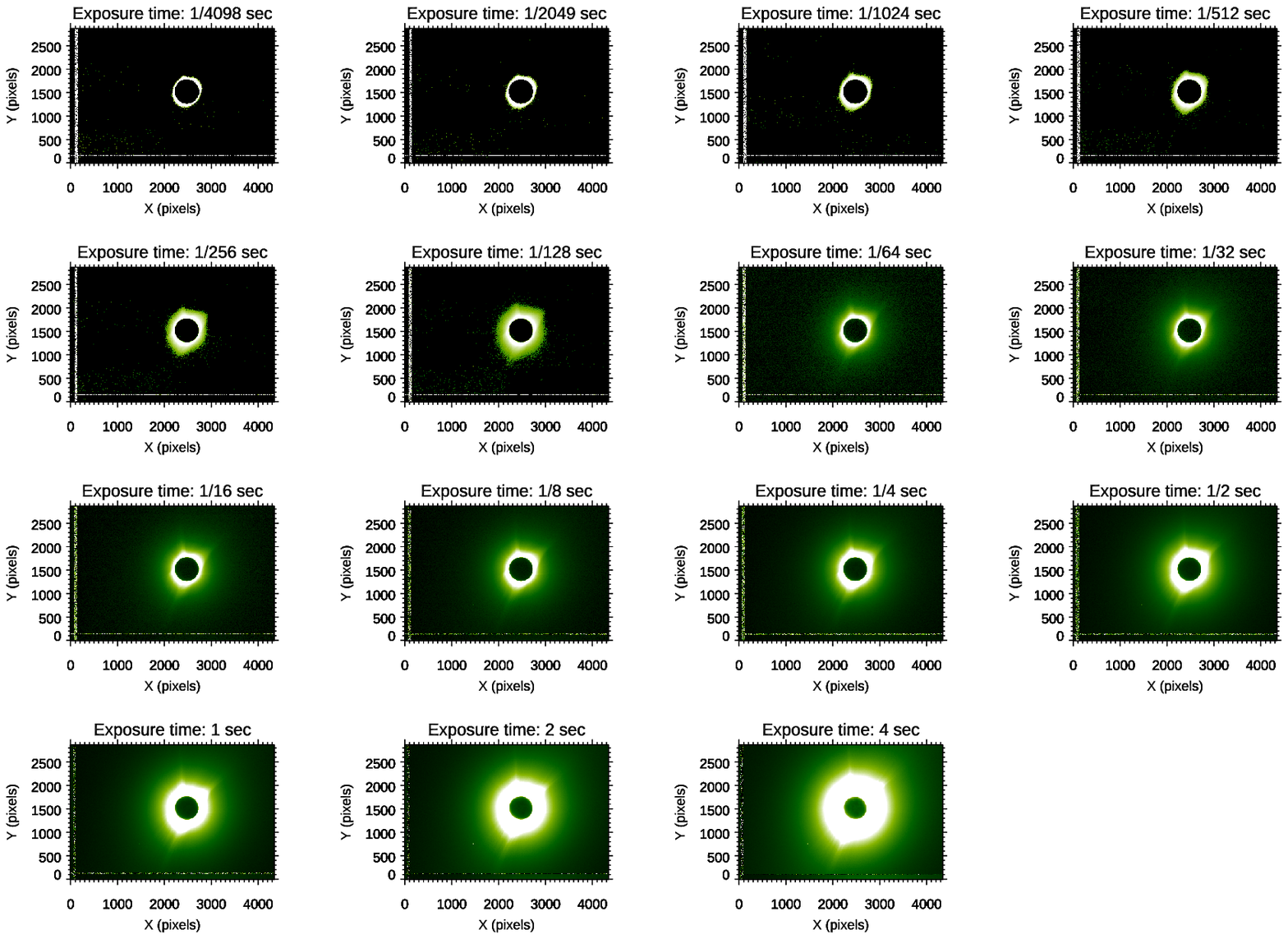}
\caption{
The first bracketing sequence of 15 exposures (G channel) acquired without the linear polarizer, after the correction for dark currents and flat field, demosaicing, co-alignement, and normalization by the exposure times.   
}
\label{Fig5}
\end{figure*}
Moreover, the 2D distributions of stellar intensities in the different colors also provide useful information on the possible optical aberrations (such as distortions and chromatic aberrations) introduced by the optical system. These effects can be quantified for the two visible stars $\alpha-$Leonis and $\nu-$Leonis, that were observed at distances from the image center (hence from the optical axis) on the order of $\sim 990$ pixels and $\sim 1480$ pixels, respectively, hence quite far from the center of the field of view. In particular, Fig.~\ref{Fig65} shows the intensity distributions of $\alpha-$Leonis and $\nu-$Leonis in different colors as obtained with single exposures acquired with exposure times by 1/64 and 1/2 sec. Results from bi-dimensional Gaussian fittings show that the 2D distributions of intensities have ratios between the Gaussian widths along the $x$ and $y$ directions $\sigma_x / \sigma_y \simeq 1.1$ for $\alpha-$Leonis and $\simeq 1.3$ for $\nu-$Leonis. Fig.~\ref{Fig65} shows that the employed optical system spreads the point light sources as if rotated about the center of the image, what is called sagittal astigmatism. On the other hand, the centroid locations of stellar emissions in the three colors (plus symbols in Fig.~\ref{Fig65}) have relative shifts by less than one pixel, leading to a limited level of chromatic aberration, shown by the right panels in Fig.~\ref{Fig65}. In summary, considering the effective image resolution given above (2.76 pixels), these aberrations will have only second-order effects, in particular for the inner corona that was observed close to center of field of view.

\subsection{Image combination} \label{sec: combination}

After image co-alignement with Regulus, it is possible to combine different exposures (once normalized by the exposure times) to get the best possible mosaic image covering the whole visible corona from the edge of the occulting disk of the Moon to larger altitudes. In particular, all the images acquired during the first bracketing sequence and after the co-alignement are shown in Fig.~\ref{Fig5} for different exposures. The image sequence shows that with shorter exposure times (top rows) only the inner corona is visible and no signal is detected farther from the Sun, while for longer exposure times (bottom rows) the outer corona becomes visible, but the inner corona is entirely saturated. This makes the combination of all images not a trivial process, as it is explained here.

\begin{figure*}[t]
\centering
\vspace{-0.5cm}
\includegraphics[width=1\textwidth]{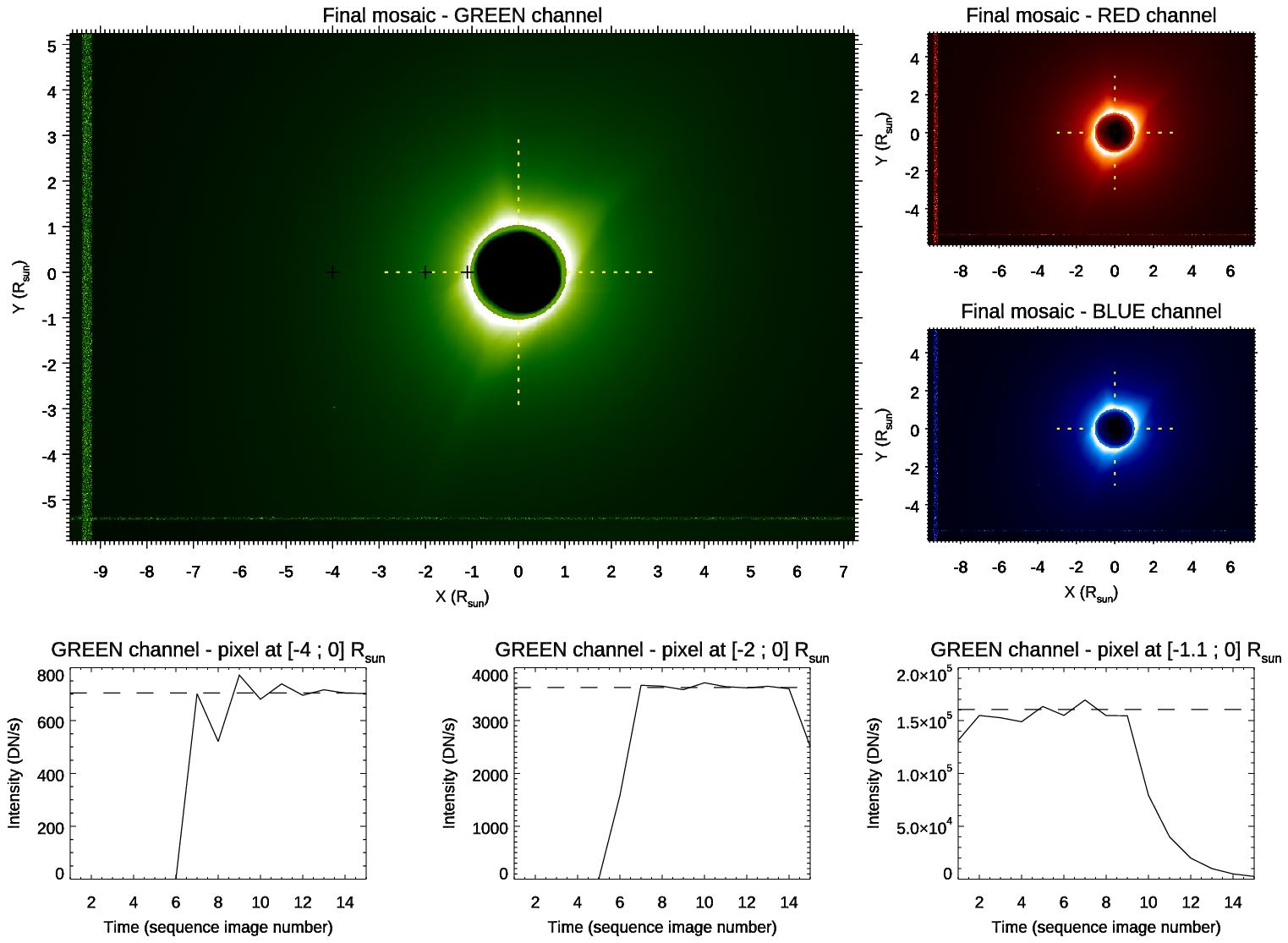}
\vspace{-1cm}
\caption{
Top: resulting mosaic image in the G channel (top left) and in the B and R channels (top right) constructed from the sequence of 15 exposures. Bottom panels: intensity (DN/s) observed in three example pixels for different exposures. The locations of the three pixels are shown by the plus symbols in the top panel. The yellow dotted lines show the locations where RGB intensity profiles have been extracted to plot panels of Fig.~\ref{Fig7}.
}
\label{Fig6}
\end{figure*}

\begin{figure*}[t!]
\centering
\includegraphics[width=0.99\textwidth]{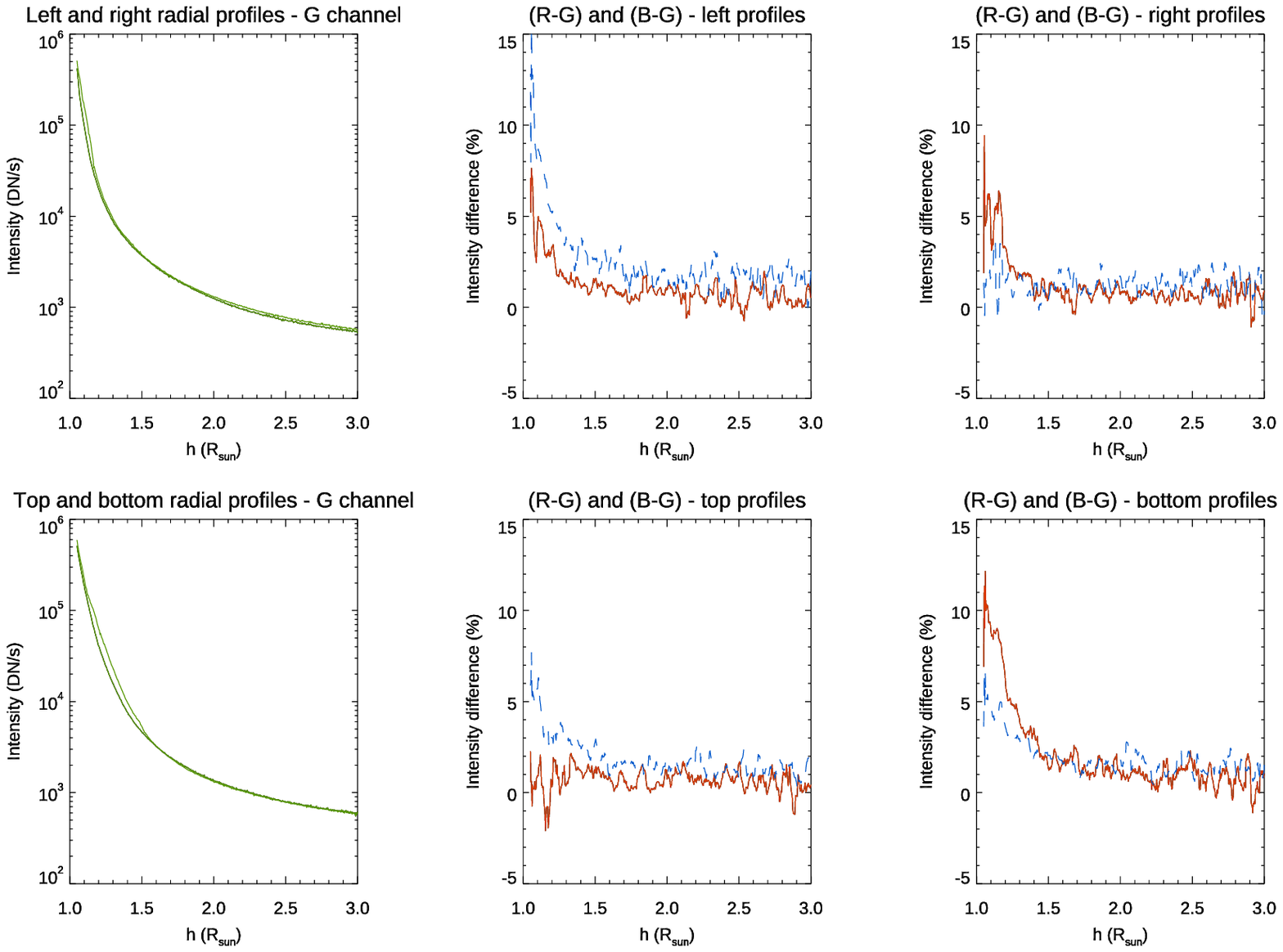}
\caption{
Left column: radial intensity (DN/s) profiles in the G channel plotted along the central row (top) and column (bottom) passing through the solar disk center as a function of heliocentric distance (yellow dotted lines in ~\ref{Fig6}). No smoothing has been applied to these curves. Middle and right columns: relative differences (\%) between the R and G (solid red lines) and between the B and G (solid blue lines) intensity profiles extracted left (top middle panel), right (top right panel), top (bottom middle panel), and bottom (bottom right panel) with respect to the solar disk center (yellow dotted lines in ~\ref{Fig6}).
}
\label{Fig7}
\end{figure*}
Once the images are normalized for the exposure times, for each pixel it is possible to plot the measured intensity for increasing exposure time. Pixels located in the inner corona have almost the same signal for shorter exposures (thanks to the linearity already shown with Regulus), while as the exposure time increases the signal saturates, and this means that after normalization for the exposure time the observed signal goes almost to zero. This behaviour is clearly shown in the bottom right panel of Fig.~\ref{Fig6}, showing the intensity (normalized to the exposure time) from different images in one the pixel located at a projected heliocentric distance of 1.1 R$_{sun}$. On the other hand, pixels located in the outer corona have almost a negligible signal in the first images, and then the signal rises becoming almost constant for longer exposure times in the linearity interval. This opposite behaviour is shown in the bottom left panel of Fig.~\ref{Fig6}, relative to the intensity measured in a pixel located at 4 R$_{sun}$. In pixels at intermediate altitudes the intensity first rises, reaching almost a constant value, and then decreases (Fig.~\ref{Fig6}, bottom middle panel relative to a pixel located at 2 R$_{sun}$).

For these reasons, the combination of the acquired images has been performed by deriving for each pixel the average intensity measured over the maximum number of exposures in the interval of linearity, determined with linear fitting. The resulting measured intensities for the three example pixels are shown by dashed horizontal lines in bottom panels of Fig.~\ref{Fig6}. The final combined image has been then constructed by iterating over all pixels and by replacing in each pixel the average intensity measured in this way. The output mosaic image for the G channel is shown in the top panel of Fig.~\ref{Fig6}; the same operations have been repeated also for the R and B channels, building two mosaics with half resolution with respect to the mosaic in the G channel. Notice that curves shown in the bottom panels of Fig.~\ref{Fig6} show again (much better than the bottom panel of Fig.~\ref{Fig4}) the good linearity of the detector response for varying exposure times.

\begin{figure*}[t!]
\centering
\vspace{-2cm}
\includegraphics[width=0.99\textwidth]{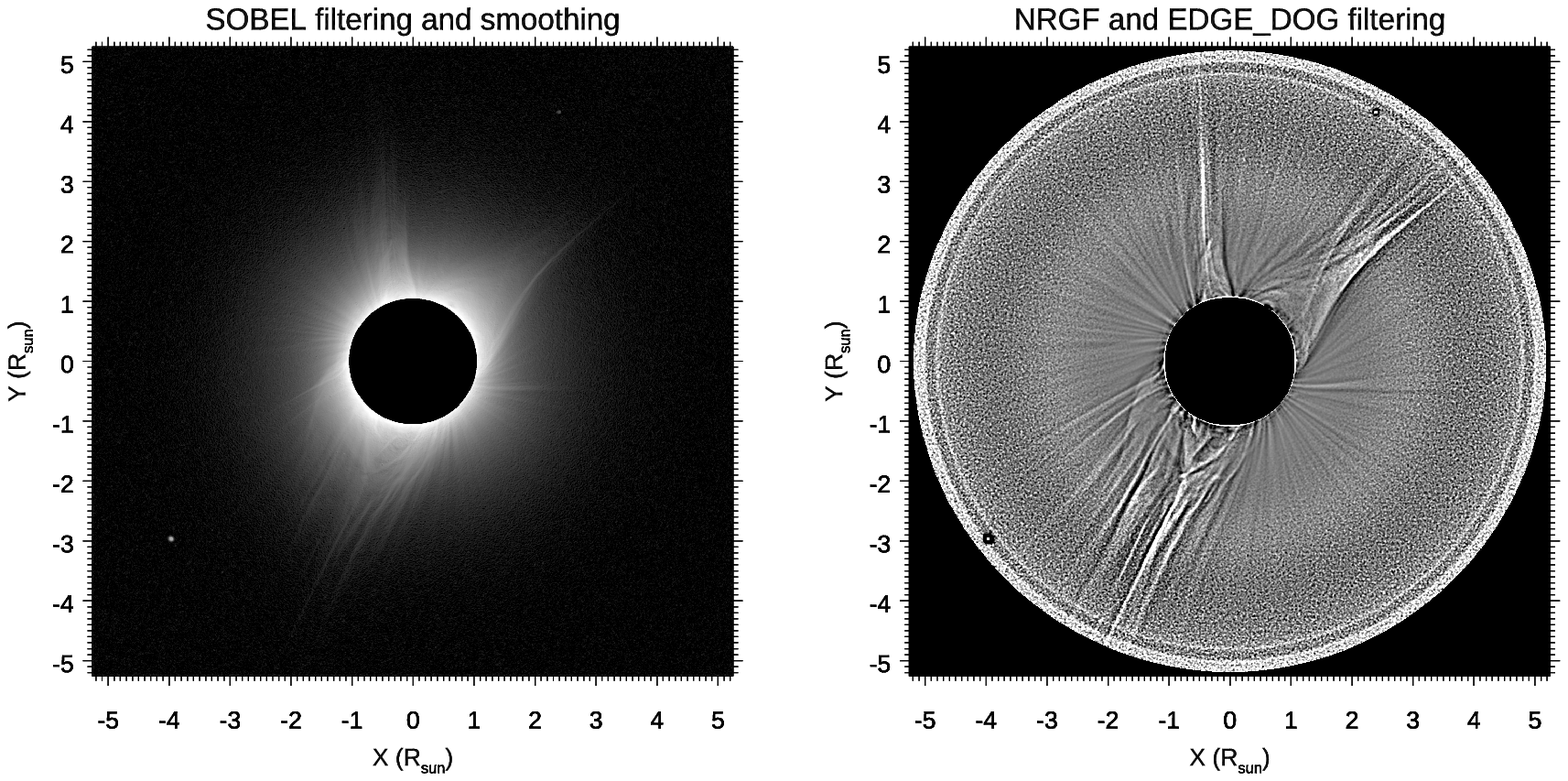}
\vspace{-1.5cm}
\caption{
Left: image resulting from the mosaic in the G channel enhanced in contrast after application of SOBEL filtering method and smoothing. Right: image resulting from the mosaic in the G channel enhanced in contrast after application of the NRGF and EDGE\_DOG filtering methods. Both images show also the $\alpha-$Leonis (close to bottom left corner) and $\nu-$Leonis (close to top right corner) stars.
}
\label{Fig8}
\end{figure*}
The same techniques described above have been applied also to co-align and combine all the exposures acquired with the linear polarizer. The only difference was that (because with the polarizer the star intensity was reduced by about a factor 4) the identification of Regulus star position in shorter frames was harder, and was made again by back-extrapolating the position of the star derived in longer exposures. Also, among the three sequences acquired with three different orientations of the polarizer, images acquired with the third and last orientation (after rotation by $\sim 45^\circ$) have not been analyzed in this work, because (due to the short duration of this TSE) as the acquisition of the third polarized sequence started, the illumination coming from the first fraction of solar disk emerging behind the edge of the Moon at the end of the TSE affected a significant part of those images \citep[see the full image sequence thumbnails showed by][]{bemporad2017}. Possible analysis of these images for limited coronal regions will be investigated in the future. 

In the end, all the above operations provided in output one mosaic image for each one of the three RGB channels, and for each one of the first three image sequences: the first one (acquired without the polarizer), and the second and third ones (acquired with two different orientations of the polarizer separated by $90^\circ$). The scientific analysis of these images is described in the next sections.

\section{Analysis of RGB mosaics} \label{sec: RGBanalysis}

\subsection{RGB intensity distributions}

Given the three mosaics in RGB channels, one of the first interesting things to do is a direct comparison between the observed radial intensity profiles in the three channels. To this end, the higher resolution G image has been interpolated to half of its resolution to become directly comparable with the R and B images. The comparison has been performed along four radial profiles, covering the projected heliocentric distances between 1.05 and 3 R$_{sun}$, and extracted along the central column (above - top - and below - bottom - the solar disk) and the central row (right and left with respect to the solar disk) passing through the disk center. The projected locations of pixels from which these profiles have been extracted are shown by yellow dotted lines in Fig.~\ref{Fig6}; the resulting curves are given in Fig.~\ref{Fig7}.

In particular, left panels in this Figure shows the intensity profiles (DN/s) in the G channel extracted left and right (Fig.~\ref{Fig7}, top left panel) and top and bottom (Fig.~\ref{Fig7}, bottom left panel) with respect to the disk center. Notice that no smoothing has been applied to these curves that have been simply extracted along single columns and rows in the resulting image mosaic: the low level of fluctuations shows the very good quality (i.e. good signal-to-noise ratio) obtained in this mosaic up to larger distances from the Sun. The other four panels in this Figure show the relative difference (\%) between R and G intensities (red solid lines) and between B and G intensities (blue solid lines) extracted left (top middle plot), right (top right plot), top (bottom middle plot), and bottom (bottom right plot) with respect to the disk center. These curves show clearly that in the inner corona (below $\sim 1.5$ R$_{sun}$) the intensities observed in the B channel are systematically higher than those in the R channel, but only in the left and top profiles, while in the right and and bottom profiles the opposite occurs, with intensities in the R channel being systematically higher than those measured in the B channel. This means that the explanation for these differences cannot be ascribed to different intensities in the band-passes of RGB filters, otherwise one should expect approximately the same behaviour in the inner corona regardless of the latitudinal location in the corona. Also, this effect cannot be ascribed to chromatic aberrations, which would have the opposite effect of increasing difference between R and B intensities going farther from the center of the field of view, hence farther from the solar disk center.

Another explanation of this effect requires to go back to the schematic representation given in Fig.~\ref{Fig2} (panel b); this Figure clearly shows that, for instance in the hypothesis of an intensity decreasing uniformly only along image rows and from right to left (as it occurs mainly in the intensity profiles extracted left with respect to the solar disk), because in each RGB quadruplet the R pixels are always located at higher altitudes with respect to the B pixels, the reconstructed intensity profiles in the B channel are systematically higher than those reconstructed in the R channel. The opposite occurs considering what I called right profiles, where the R pixels are always located at lower altitudes with respect to the B pixels; a similar difference is present in the top profiles with respect to the bottom profiles. In summary, this means that the observed intensity differences between RGB pixels are mostly due to the different locations of those pixels. In particular, in the inner corona (below $\sim 1.5$ R$_{sun}$) the intensity radial gradients are so high, than even the small difference in the projected altitude of nearby pixels (separated by 3.7 arcsecs, corresponding to $3.9 \times 10^{-3}$ R$_{sun}$) results in considerable relative intensity differences (up to $\sim 5-10\%$).

This result has important implications that are discussed here. Over the last few years different authors demonstrated the advantages to use what is called a PolarCam or Polarization Imaging Camera to observe the solar corona \citep{reginald2017, burkepile2017, gopalswamyashiro2018, reginald2019, judge2019, fineschi2019, vorobiev2020}. This instrument consists in a camera (or a telescope, or any other optical system) equipped with a sensor having a micro-polarizer array placed over the sensors (pixels) with four alternating orientations of linear polarizers. In practice, this is conceptually similar to a DSLR camera, where the 4 RGB filters of the Bayer matrix (Fig.~\ref{Fig2}, top left) are replaced with 4 linear micro-polarizers. The obvious advantage of this camera is that the acquisition of a single exposure is sufficient to have in each super-pixel 4 different measurements of the linear polarization at 4 different angles (0, 45, 90 and 135 degrees), allowing the optimal measurement of the total and polarized brightness of the corona and also high-cadence observations. Recently the use of similar cameras has been also proposed as a payload of small satellites for solar coronagraphy from space \citep{gopalswamy2018}.

Nevertheless, the analysis of images acquired by a PolarCam is usually performed by assuming that the 4 sub-pixels in the same macropixel are sampling the same coronal plasma with 4 different orientations of the polarizer, then reconstructing the 4 images corresponding to different polarizer orientations by simply collecting together nearby pixels having the same micro-polarizer orientation \citep[e.g. Fig. 2 by][]{reginald2017}, without any interpolation to a common spatial grid. This working hypothesis is not always applicable, because each pixel is illuminated by a different coronal region, in principle. The problem is partially mitigated by the fact that the effective resolution is broadened by the instrument PSF and - for ground-based observations - also by the astronomical seeing. Nevertheless, Fig.~\ref{Fig7} shows that, even if these observations were acquired with a pixel projected size by 3.7 arcsec/pixel and an effective image resolution by 2.76 pixels, the intensity gradients in the inner corona are so steep, that the different locations of RGB pixels result in different illumination levels. For a direct comparison, the PolarCam projected pixel sizes employed by \citet{reginald2017}, \citet[][]{judge2019} and \citet{fineschi2019} were 3.3 arcsec, 2.87 arcsecs, and 4.3 arcsec respectively, hence comparable or even larger than projected pixel size of the DSLR camera employed in this work.

Thus, the main consequence of plots shown in Fig.~\ref{Fig7} is that careful must be given in the analysis of data acquired with PolarCams in the inner regions of the solar corona (i.e. below 1.5 R$_{sun}$): an error by up to $\sim 10 \%$ could be present in the derived measurements of total and polarized brightness, hence in the electron density measurements.

\subsection{Image filtering}

One of the most interesting advantages of TSE observations is the possibility to observe not only large scale, but also smaller scale and fainter features as density inhomegeneities expanding from very close to the solar limb. The observation of these features is very important, in particular because their orientation is usually believed to match the orientation of the coronal magnetic fields \citep[see e.g.][]{boe2020}, that are dominating the dynamic of coronal plasma, but cannot be easily measured. In order to enhance the visibility of fainter coronal features in eclipse images, usually filtering methods are required to flatten the strong radial intensity variations on the one hand, and also to increase the image contrast on the other hand. Many methods have been developed for these purposes over the last decades by different authors, such as the multidirectional maximum of second derivatives method \citep{koutchmy1988}, the Adaptive Circular High-pass Filter (ACHF) method \citep{druckmuller2006}, the normalizing-radial-graded filter (NRGF) method \citep{morgan2006}, the application of high-pass filters to improve signal-to-noise ratio \citep{deforest2018}, and many other methods \citep[see reviews by][and references therein]{pasachoff2007, rusin2020}.
\begin{figure*}[t!]
\centering
\vspace{-1.5cm}
\includegraphics[width=0.98\textwidth]{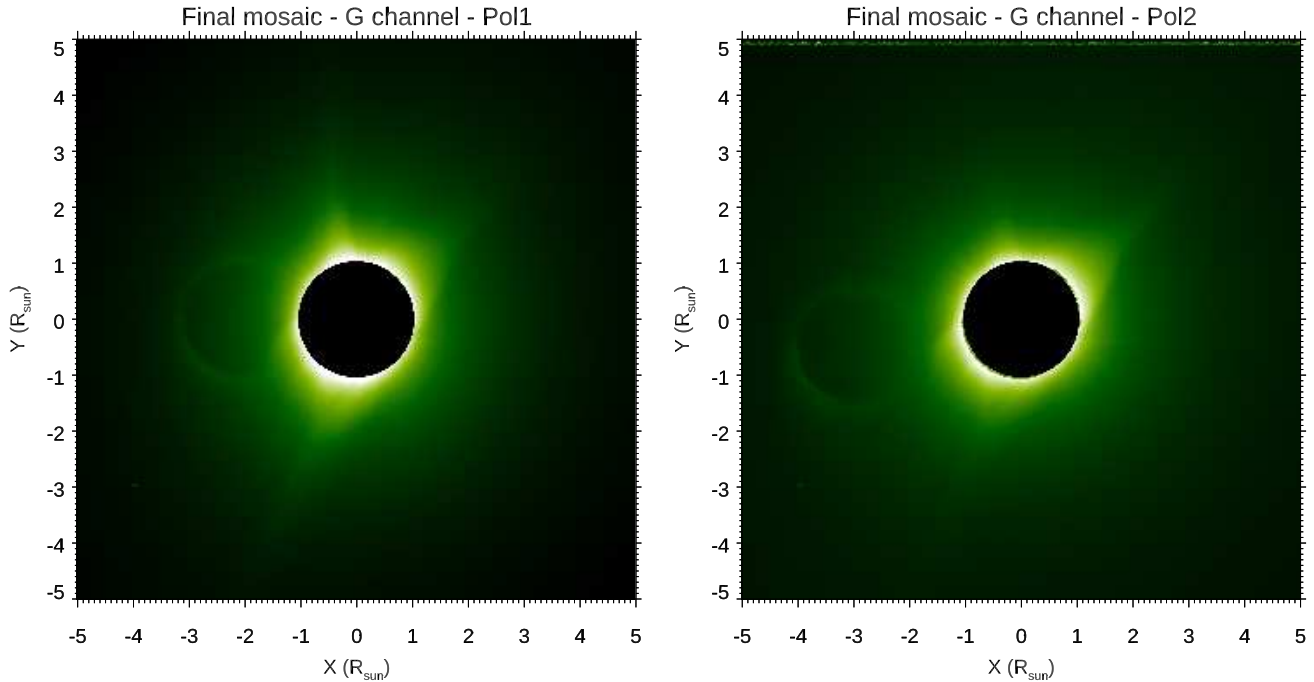}
\vspace{-1.5cm}
\caption{
Resulting mosaic images in the G channel constructed from the two sequences of 15 exposures acquired with the first (left) and the second (right) orientations of the linear polarizer. Both images show the presence of a ghost due to internal reflections by the linear polarizing filter.
}
\label{Fig9}
\end{figure*}

\begin{figure*}[t!]
\centering
\includegraphics[width=0.99\textwidth]{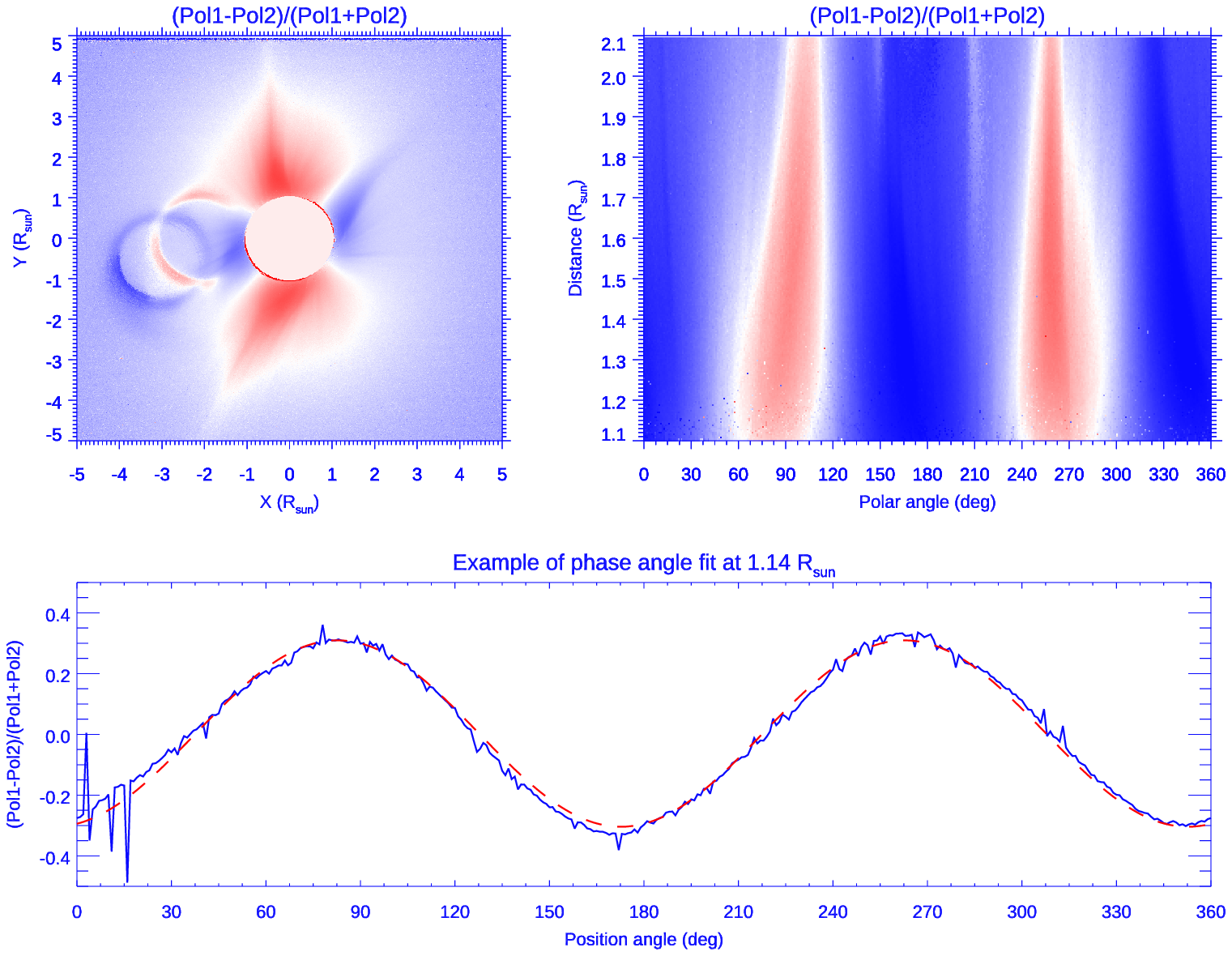}
\caption{
Top left: relative difference between the two mosaics in the G channel (Pol1 and Pol2) acquired with two different orientations of the linear polarizer separated by 90$^\circ$. The color scale shows positive (negative) values plotted as red (blue) colors. Top right: the same image transformed in polar coordinates and showing the modulation of polarization between 1.1 and 2.1 R$_{sun}$. Bottom: the latitudinal variation of intensity at constant altitude (solid line) and the corresponding sinusoidal fit (dashed line), plotted as a function of the position angle (running counter-clockwise from $X-$axis).
}
\label{Fig10}
\end{figure*}
In this work two simple combinations of standard mathematical methods were applied, methods that are freely available and distributed. The first one consists first in the application of the SOBEL filter (an edge enhancement operator based on the detection of maximum image gradient directions), followed by a simple image smoothing to reduce the noise. The resulting image (Fig.~\ref{Fig8}, left panel) plotted in Log scale is quite similar to the natural appearance to the human naked eye of the solar corona during TSE, and shows not only the orientation of main coronal features (such as coronal streamer and plumes), but also the location of the two visible stars: Regulus or $\alpha-$Leonis (close to the bottom left corner), and $\nu-$Leonis (close to the top right corner). 

The second combination consists in the application of NRGF filter (freely distributed under SolarSoftware), followed by the so-called EDGE\_DOG filter (a band-pass filter based on the subtraction of two copies of the same image obtained after the application of different Gaussian blurrings). The resulting image (Fig.~\ref{Fig8}, right panel) plotted in linear scale shows much better the outward extension of the fainter coronal features (plumes), that can be followed up to $\sim 3$ R$_{sun}$, while brighter features (streamers) are visible up to the image edge located at $\sim 5$ R$_{sun}$; the $\alpha-$Leonis and $\nu-$Leonis stars are also clearly visible.

The filtered images can be used also to co-align the TSE observations with the solar North, based on images provided by other ground-based or space-based observatories. The determination of the rotation angle was performed here based on the calibrated total brightness image derived from the analysis of the polarized sequence, which is discussed below. Moreover, in this work the filtered images have been employed to derive the position of the center of the Sun behind the occulting lunar disk, given by the measured position of the visible stars and by the celestial coordinates of these stars and the Sun during TSE.

\section{Analysis of polarized sequences} \label{sec: POLanalysis}

The mosaic images resulting from the combination of the two sequences of 15 exposures acquired with two different orientations of the linear polarizer (called Pol1 and Pol2, and separated by 90$^\circ$) are shown in Fig.~\ref{Fig9}. The two images show, unfortunately, the presence of a ghost (due to unavoidable internal reflections from the linear polarizing filter); very similar features were present for instance also in images acquired by \cite{snik2020} (see their Fig. 2). Because it is in principle unknown how the internal reflections from the filter modified the fraction of polarized emission from the corona, removal of these artifacts was not performed. In any case, these artifacts in the images affected only small limited regions of the observed corona, and their presence is clearly identifiable. 

\begin{figure*}[t!]
\centering
\includegraphics[width=0.97\textwidth]{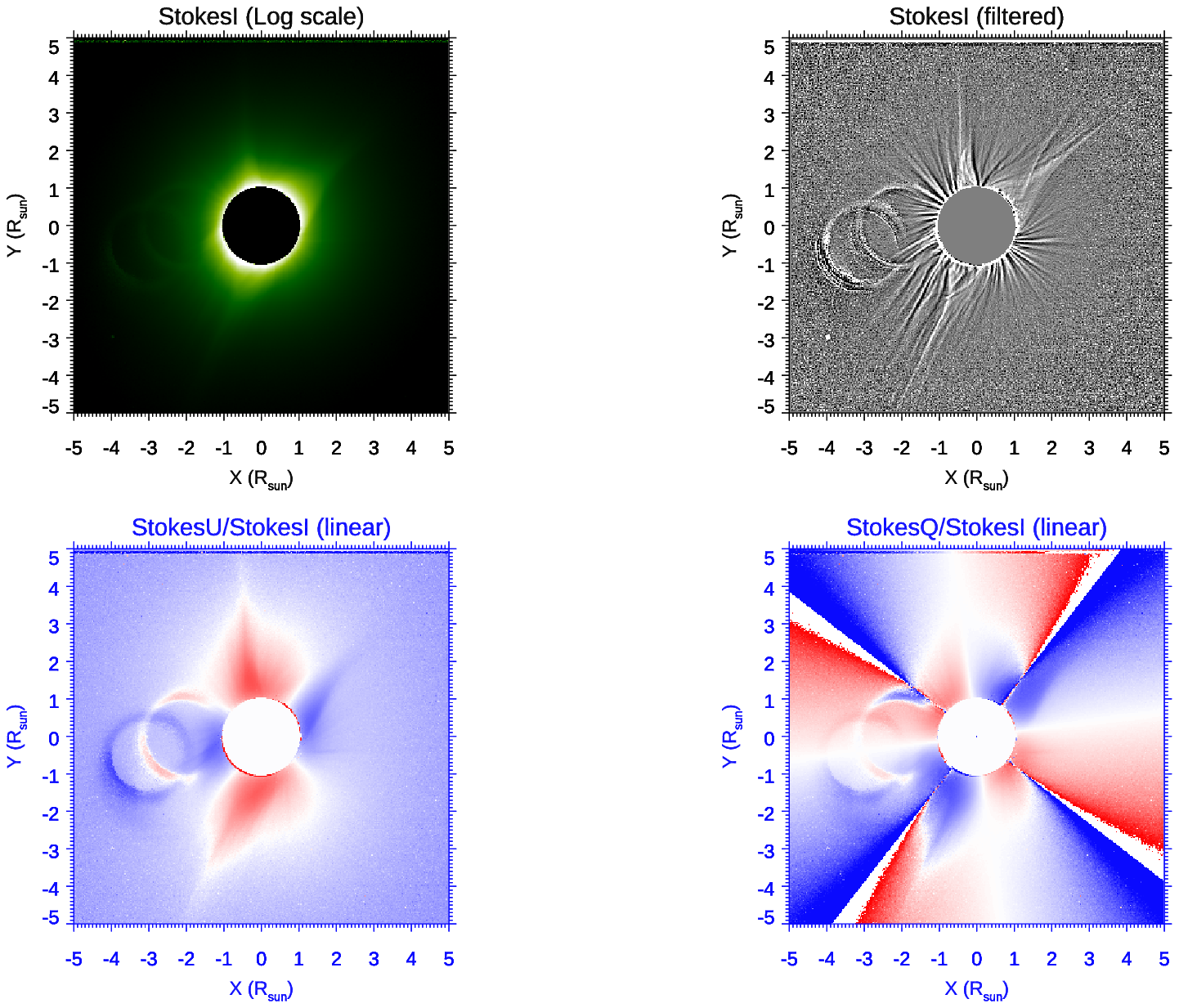}
\caption{
Top: 2D distribution of the Stokes vector $I$ component plotted in Log scale (left panel) and in linear scale after filtering (right panel). Bottom: the corresponding 2D distributions of the Stokes vector $U$ (left panel) and $Q$ (right panel) components plotted in linear scale after normalization over the $I$ component. The top right image has been filtered by simply subtracting from the original image in Log scale a median image obtained by replacing in each pixel with the average value over the surrounding area by $(0.1 \times 0.1)$ R$_{sun}$.
}
\label{Fig11}
\end{figure*}
The relative difference between these two images (Fig.~\ref{Fig10}, top left panel) clearly shows the so-called polarization cross, which is the cross-shaped distribution of pixels having almost a zero signal in the difference between Pol1 and Pol2 images (plotted with white color in Fig.~\ref{Fig10}, top left panel). The well known existence of this polarization cross is related with the specific characteristics of the coronal emission which is linearly polarized. In general, the linear polarization is described by introducing the $I$, $Q$ and $U$ components of the Stokes vector $S = [I,Q,U]$ representing the linearly polarized emission in each pixel $(i,j)$. In the simple case of no ellipticity of polarization (hence no circular polarization), the $Q$ and $U$ components of the Stokes vector are given by
\begin{eqnarray}
    Q = I\,p\,\cos(2\alpha) \\
    U = I\,p\,\sin(2\alpha) \
\end{eqnarray}
where $I$ [DN/s] is the total intensity, $p = \sqrt{Q^2+U^2}/I$ [\%] is the degree of linear polarization, and $\alpha$ is the angle of linear polarization, representing the angle of the direction of electric field oscillation from a given plane. In the specific case of the solar K-corona emission (due to Thomson scattering of photospheric emission from free coronal electrons), the radiation have a partial linear polarization, and the orientation of the linear polarization vector is always tangent to the solar limb. This means that, for the position angles where $\sin(2\alpha) = \cos(2\alpha)$, the $Q$ and $U$ components are equal, and this happens for $2 \alpha = 45^\circ + k 90^\circ$ ($k=0,1,2,3$). Given the 2D intensity distributions $I_{pol1}(i,j)$ and $I_{pol2}(i,j)$ of the two images Pol1 and Pol2 acquired with the two different orientations of a linear polarizing filter separated by 90$^\circ$, the distributions $I(i,j)$, $Q(i,j)$ and $U(i,j)$ of the Stokes vector components are given by \citep[see e.g.][]{reginald2017}:
\begin{eqnarray}
    I(i,j) & = & I_{pol2}(i,j) + I_{pol1}(i,j) \\
    U(i,j) & = & I_{pol2}(i,j) - I_{pol1}(i,j) \\
    Q(i,j) & = & U(i,j)/\tan 2 \alpha(i,j). \
\end{eqnarray}
Hence, the difference between Pol1 and Pol2 images provides directly the $U$ component of the Stokes vector, and this difference is zero around angles separated by $90^\circ$, leading to the polarization cross pattern shown in Fig.~\ref{Fig10} (top left panel).

\subsection{Determination of the phase angle}

Because (as mentioned above) the linear polarization vector in the solar corona is always tangent to the solar limb, the 2D distribution $(i,j)$ of the angle $\alpha(i,j)$ in the acquired images is easily determined, once the position of the center of the Sun is well know. Nevertheless, the angle $\alpha$ has to be corrected for a phase angle $\phi$, related with the reference angular orientation of the linear polarizer for instance with respect to the columns of images, which is in general unknown. In order to measure $\phi$, the normalized $U_{norm}$ intensity image $U_{norm} = (I_{pol2} - I_{pol1})/(I_{pol2} + I_{pol1})$ [\%] has been converted in polar coordinates, as shown in Fig.~\ref{Fig10} (top right panel). The latitudinal distributions of $U_{norm}$ have been extracted only in the inner coronal regions (where the signal-to-noise ratio is higher) and fitted with a sinusoidal function, as it is shown in Fig.~\ref{Fig10} (bottom panel). The resulting value of $\phi = -75.3^\circ \pm 0.3^\circ$ has been employed then to derive the 2D distribution of the angle $2\alpha(i,j) = 2\arctan Y(i,j)/X(i,j) +\phi$, where $X$ and $Y$ coordinates are standard Cartesian coordinates in reference system centered on the Sun.
\begin{figure*}[t!]
\centering
\vspace{-1cm}
\includegraphics[width=0.95\textwidth]{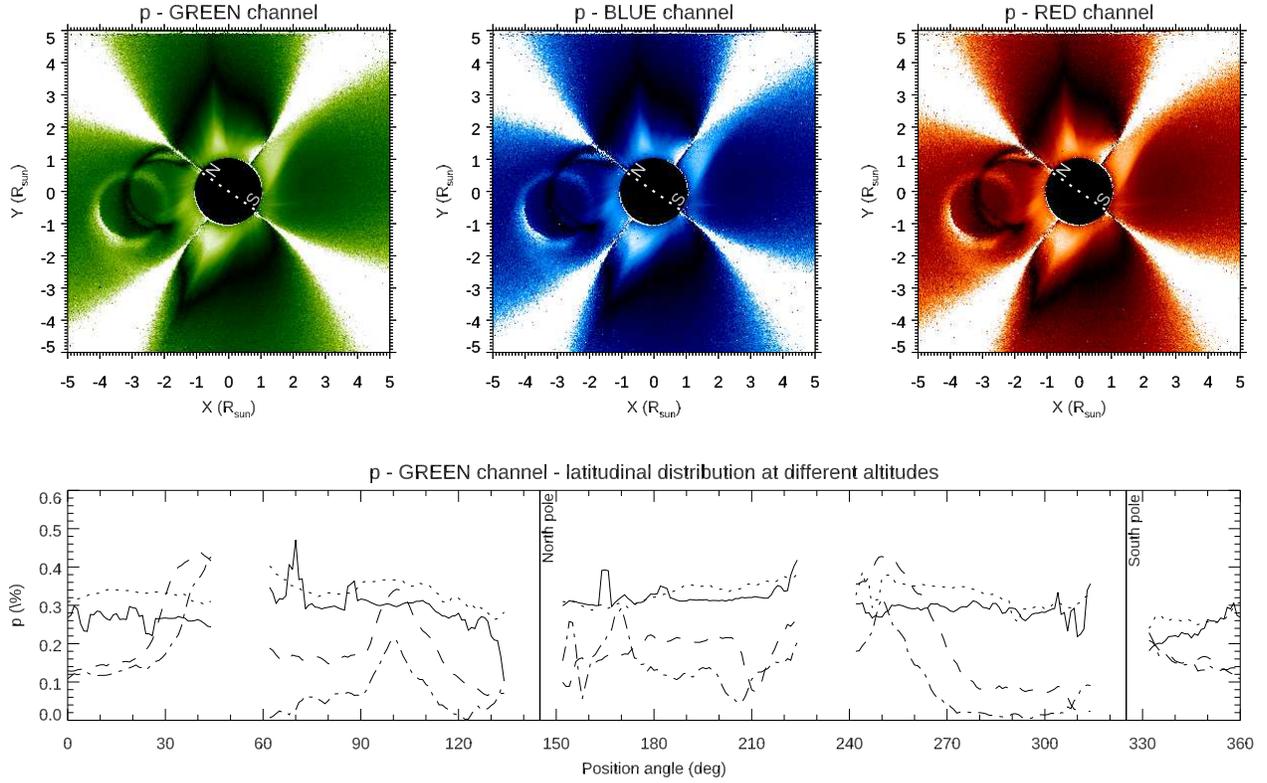}
\caption{
Top panels: degree of linear polarization $p$ as derived from the green (left), blue (middle), and red (right) channels (linear color scales going from 0 to 0.5). The $p$ distribution is affected by artifacts related to the reflection ghosts by the polarizing filter (left in each image), and also by the cross-like divergent pattern for angles where $2 \alpha = k 90^\circ$ (see text). The dotted line marks the polar axis of the Sun. Bottom: latitudinal distribution of $p$ (counter-clockwise from $X-$axis) plotted at 1.1 R$_{sun}$ (solid line), 1.2 R$_{sun}$ (dotted line), 2 R$_{sun}$ (dashed line), and 2.5 R$_{sun}$ (dash-dotted line); vertical lines marks the poles of the Sun. Values around the cross-like divergent pattern have been omitted in these plots.
}
\label{Fig12}
\end{figure*}

\subsection{Determination of the degree of linear polarization}

Once the 2D distribution of $\alpha$ angle is determined, Eqs. (1-5) provide directly a measurement of the Stokes vector components $[I, Q, U]$, as well as the degree of linear polarization $p$; the resulting 2D distributions of $[I, Q, U]$ components are shown in Fig.~\ref{Fig11}. The main problem with the analysis described here is that it is based on the combination only of two different images (Fig.~\ref{Fig9}) acquired with two different orientations of the linear polarizer. This results in the relatively simple Eqs. (3-5), but these equations have two main disadvantages. First, the 2D distribution of angle $\alpha$ has been determined with respect to the position of the center of the Sun, which is in general not known. As mentioned, for this TSE campaign, thanks to the presence of stars, the center of the Sun behind the lunar disk has been determined in the mosaic images, but this is not always possible. Second, having only two polarized images, the expression for the $Q$ component (Eq. 5) has a tangent function in the denominator, and for pixels close to the angular positions where $\alpha = k \, 90^\circ$ ($k=0,1,2,3$) the expression for $Q$ diverges, leading to unreliable values in the cross-like pattern visible in Fig.~\ref{Fig11} (bottom right panel). 

For these reasons, measurements of $p$ [\%] in the solar corona are usually performed by acquiring at least 3 images with 3 different orientations of the linear polarizer. Unfortunately, as explained above (end of Section \ref{sec: combination}), the third and last sequence of polarized images acquired during this TSE is affected by light coming from the solar disk emerging behind the Moon at the end of totality, and these images have not been analyzed here. In any case, despite the above problems related with the use of only two polarized mosaic images, the 2D distribution of degree of linear polarization $p$ was successfully determined, and this was done independently for the three RGB channels. The resulting $p$ images (Fig.~\ref{Fig12}, top panels) are affected by artifacts related with the reflection ghosts (Fig.~\ref{Fig9}), and also by the mentioned cross-like pattern of diverging pixels. The resulting 2D distribution of $p$ [\%] appears to be well determined, at least in the inner coronal regions, and in good agreement with values provided for instance by \cite{snik2020} (see their Fig. 2). The orientation of the polar axis of the Sun in these images has been determined from a co-alignement with images acquired by the Mauna Loa COSMO K-Coronagraph (see below).
\begin{figure*}[t!]
\centering
\vspace{-2.2cm}
\includegraphics[width=0.99\textwidth]{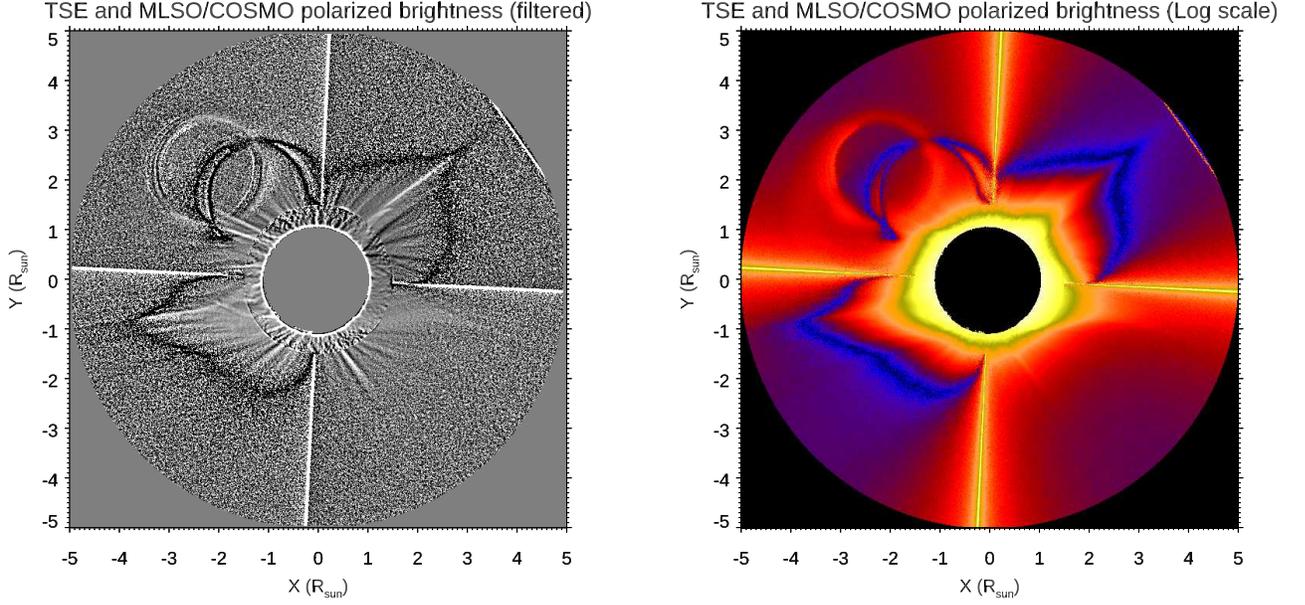}
\vspace{-2.cm}
\caption{
Combined images obtained from a superposition of $pB$ values measured by MLSO/COSMO (shown in the range between 1 and 1.5 R$_{sun}$) and those measured in this work with TSE observations (shown in the range between 1.5 and 5 R$_{sun}$), after co-alignement, rotation, and inter-calibration with MSLO/COSMO values. The right image shows the $pB$ [1/B$_{sun}$] plotted in Log scale, while the left image shows the nice correspondence between fainter coronal features after image filtering.  
}
\label{Fig13}
\end{figure*}

In what follows I describe how the resulting $p$ image in the G channel has been radiometrically calibrated to measure the polarized brightness $pB$ and further analyzed to measure the 2D distribution of coronal electron densities.

\section{Image radiometric calibrations} \label{sec: radiometric}

In order to derive the electron densities $n_e$ [cm$^{-3}$] from the observed coronal emission, it is necessary to convert the intensities into physical units, typically in mean solar brightness (MSB); this corresponds to perform the radiometric calibration of images. In this work the radiometric calibration has been performed first with respect to another instrument providing calibrated intensities (relative calibration), and second by employing the observed total solar brightness and the brightness of visible stars (absolute calibrations).

\subsection{Relative radiometric calibration}

The relative radiometric calibration has been performed by re-scaling the values of polarized brightness $pB = I\,p$ [DN/s] derived here to those measured by the Mauna Loa COSMO K-Coronagraph (K-Cor) in Hawaii [1/B$_{sun}$]. This telescope provides $pB$ images in a field of view from 1.05 to 3 R$_{sun}$ with 5.64 arcsec/pixels and a spatial resolution of 11.29 arcsec. For the inter-calibration I employed in particular the 2 min averaged MLSO image at 17:43:55 UT. Once the TSE and MLSO images are coaligned with the Sun center, the rotation angle to be applied to TSE images has been determined from a comparison between the latitudinal location of fainter coronal features visible in filtered images (Fig.~\ref{Fig13}, left panel). Then, the comparison between $pB$ values provided the calibration factor K$_{G}$ to convert $pB$ measurements from TSE images (in units of DN/s) to $pB$ measurements from MLSO (in units of 1/B$_{sun}$); the resulting value of the calibration factor is K$_G = 1.54 \times 10^{11}$ B$_{sun}$ DN/s. 

\begin{figure*}[t!]
\centering
\includegraphics[width=0.95\textwidth]{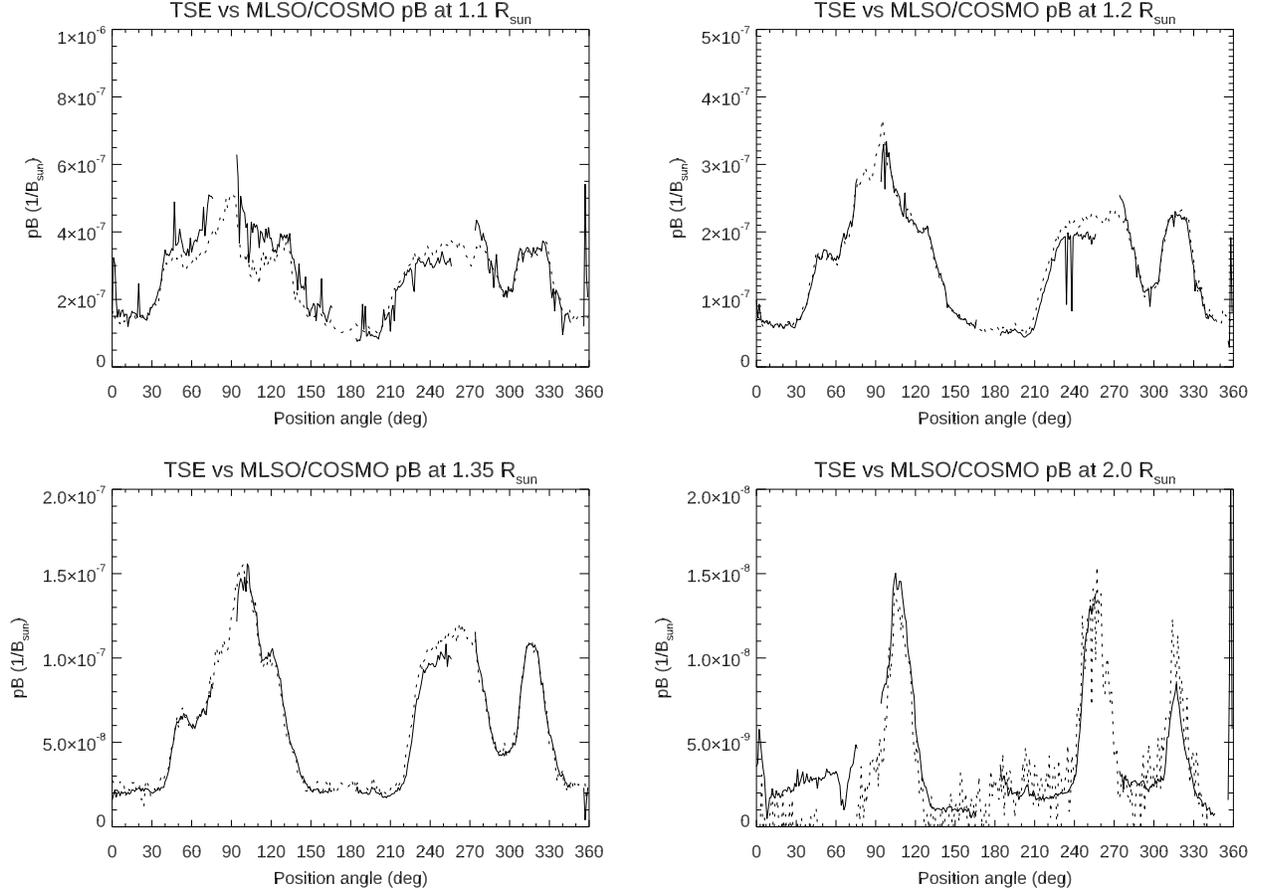}
\caption{
Comparisons between the latitudinal distribution of $pB$ values measured here from TSE (solid lines) and provided by MLSO (dotted lines) at 1.1 R$_{sun}$ (top left), 1.2 R$_{sun}$ (top right), 1.35 R$_{sun}$ (bottom left), and 2.0 R$_{sun}$ (bottom right). Again, values around the cross-like divergent pattern have been omitted in these plots. No smoothing have applied to the data; the bottom left panel is directly comparable with Fig. 9 (bottom panel) by \cite{judge2019}.
}
\label{Fig14}
\end{figure*}
The top right image in Fig.~\ref{Fig13} shows again the presence of reflection ghosts in TSE images (top left quadrant), and the cross-like divergent pattern discussed above, but more importantly the comparison shows a really striking agreement between TSE and MLSO values after re-scaling with the above calibration factor, so that the boundary region between the two TSE and MLSO images (arbitrarily assumed at 1.5 R$_{sun}$) is not even visible. 

A more quantitative comparison between $pB$ values measured here from TSE images and provided by MLSO is shown in Fig.~\ref{Fig14}. These plots (providing the measured values without any smoothing to the data) shows a very good agreement between $pB$ values not only in their latitudinal distribution, but also in their radial variations, an agreement which is comparable for instance with what obtained by \cite{judge2019} with a much more complex instrumentation. The bottom right panel of Fig.~\ref{Fig14} also shows that at 2 R$_{sun}$ the quality of $pB$ measurements obtained with a DSLR camera during TSE have a lower noise level with respect to measurements obtained with the MLSO/COSMO coronagraph.

\subsection{Absolute radiometric calibrations}

More than just the relative radiometric calibration, the absolute radiometric calibration has also been determined here with two different methods. The first method is based on the measurement of the total brightness of the solar disk B$_{sun}$: given the full-Sun images acquired before the beginning of PSE, in principle this measurement is not difficult. The main problem is to measure the transmittance $T$ of the Baader OD5.0 solar filter that was employed for the observations before the TSE. A transmittance curve as a function of wavelength is not provided by the manufacturer of this filter, but this was measured by \cite{koukarine2013}, who provided (their Fig. 4) the measured transmittance curve. From this Figure, the transmittance values $T$ have been extracted and averaged over different wavelength intervals, obtaining a value for the transmittance in the G channel ($T_G = 5.30 \times 10^{-6}$), as well as corresponding values for the R ($T_R = 4.45 \times 10^{-6}$) and the B ($T_B = 6.11 \times 10^{-6}$) channels.

The selected full Sun image has been then analyzed by repeating exactly the same steps explained above for the TSE images, and by separating the image in three RGB images. Finally, the MSB values B$_{sun,G}$, B$_{sun,R}$, and B$_{sun,B}$ [DN/(pix$^2$ s)] have been computed for each on the three RGB colors as \citep[see][]{allen2000}
\begin{equation}
B_{sun,RGB} = F_{RGB} / (\pi\, A_{RGB}\, T_{RGB}\, t_{exp})
\end{equation}
where $F_{RGB} = I_{RGB} t_{exp}$ [DN] is the intensity flux obtained by integrating over the whole visible solar disk in each image, $A_{RGB}$ [pix$^2$] is the area covered by the solar disk in pixels, $T_{RGB}$ are the transmittances given above, and $t_{exp}$ [s] is the exposure time. Resulting values computed with the above formula are $B_{G,sun} = 3.26 \times 10^{11}$ DN/(pix$^2$ s), $B_{R,sun} = 3.56 \times 10^{11}$ DN/(pix$^2$ s), and $B_{B,sun} = 3.06 \times 10^{11}$ DN/(pix$^2$ s) respectively for the G, R, and B channels. In particular, the derived value of $B_{sun,G}$ is about a factor $\sim 2$ larger than the normalization constant K$_G$ determined with relative inter-calibration to re-scale TSE measurements of $pB$ to values provided by MLSO. The origin for this disagreement is not known, but considering the amateur equipment employed in this work, and the large uncertainties in particular in the measurement of Baader OD5.0 filter transmittance, such a disagreement can be considered as acceptable and not surprising.

The presence of visible stars in TSE images provides also an alternative method to determine the absolute radiometric calibration, based on the observed star intensities. In particular, from bi-dimensional Gaussian fitting of $\alpha-$Leonis and $\nu-$Leonis stars observed in the G channel in the last 4 exposures acquired during the first TSE sequence, it turns out that the average intensities $I_G$ [DN/s] for the two stars are $I_{G,\alpha} = (3.64 \pm 0.05) \times 10^5$ DN/s and $I_{G,\nu} = (1.1 \pm 0.1) \times 10^4$ DN/s. This corresponds to an observed magnitude difference $\Delta m_{G,obs} = -2.5 \log_{10} (I_{G,\nu} /I_{G,\alpha}) = 3.80$. This is in very good agreement with the known visual magnitude difference $\Delta m = m_\nu - m_\alpha = 5.15-1.35 = 3.80$ as provided by \href{http://stellarium.org/}{Stellarium} \citep[based on the Naval Observatory Merged Astrometric Dataset - NOMAD; see][]{zacharias2004}, including atmospheric extinction. This also suggests that intensities measured in the G channel represents in first approximation the intensities of stars in the V-band. On the other hand, the total intensity of the corona, summing in the observed sky region between $\pm 5$ R$_{sun}$ both in $X$ and $Y$ coordinates, is $I_{G,cor} = 3.77 \times 10^{10}$ DN/s, corresponding to a total magnitude of the solar corona in the G channel (close to the V-band) $m_{G,cor} = m_{G,\alpha} - 2.5 \log_{10}(I_{G,cor} / I_{G,\alpha}) = -11.19$. This can be compared with the Moon visual apparent magnitude $m_{moon} = -12.73$ \citep{allen2000}, and implies that the observed solar corona was approximately a factor $\sim 4$ dimmer that the full Moon. 
\begin{figure*}[t!]
\centering
\vspace{-0.9cm}
\includegraphics[width=0.95\textwidth]{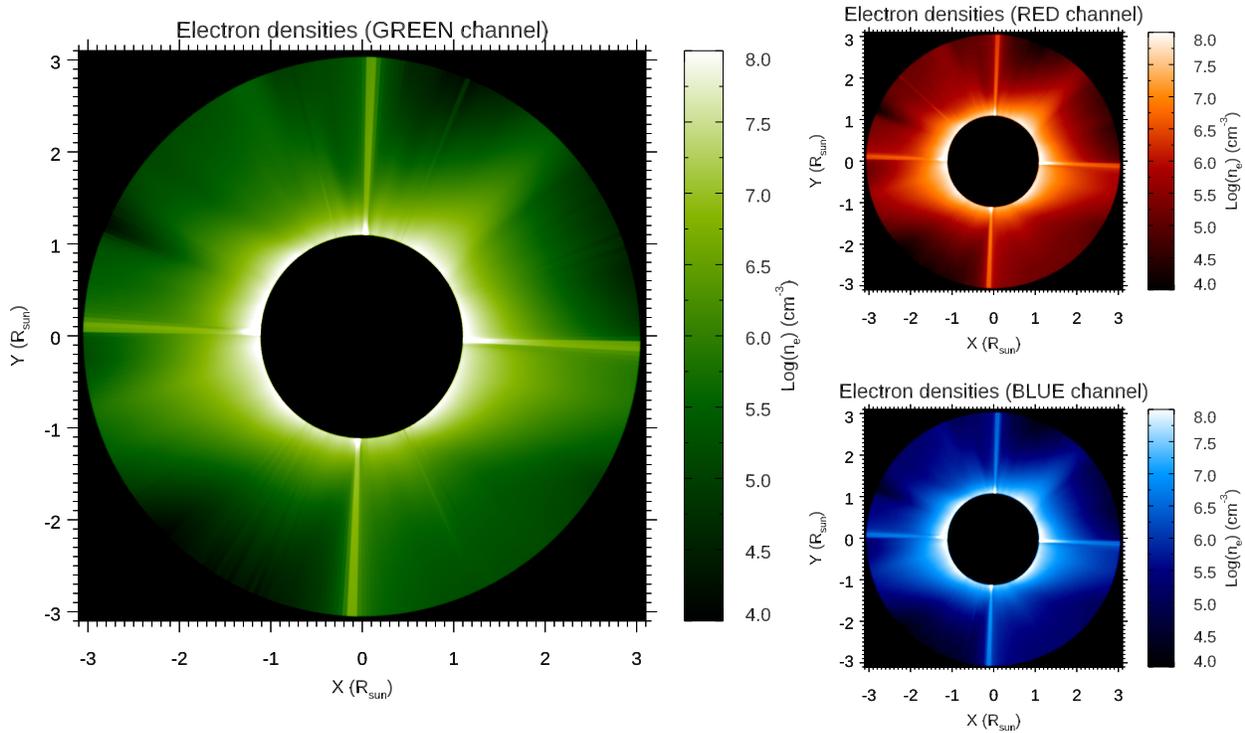}
\vspace{-0.7cm}
\caption{
Electron density maps as derived from the inversion of $pB$ maps obtained with three different pixel colors, and in particular for the green (left), red (top right), and blue (bottom right) pixels.
}
\label{Fig15}
\end{figure*}

The above measurements can be also employed to derive another estimate for the intensity of the full Sun $I_{G,sun}$ [DN/s], hence for the value of MSB, thus providing another method for the absolute calibration. Starting from the known apparent visual magnitude of the Sun $m_{sun} = -26.75$ \citep{allen2000}, and given the above total intensity and magnitude of the corona, the corresponding intensity [DN/s] of the full Sun is given by
\begin{equation}
I_{G,sun} = I_{G,cor} \cdot 10^{(m_{G,cor} - m_{G,sun})/2.5}
\end{equation}
that can be converted in $B_{G,sun} = I_{G,sun}/A_G = 2.99 \times 10^{11}$ DN/(pix$^2$ s). In the above estimates it is assumed (as suggested by the agreement between the known visual magnitude difference and observed intensity ratio for $\alpha-$Leo and $\nu-$Leo in the G channel) that the known visual magnitudes for stars and the Sun correspond with the intensity fluxes measured here from the G channel. The above value is quite close to the MSB value measured independently with the full Sun disk image $B_{G,sun} = 3.26 \times 10^{11}$ DN/(pix$^2$ s), and again approximately a factor $\sim 2$ larger than the normalization constant needed to re-scale TSE measurements to values provided by MLSO. The reason for this disagreement is not known, but for this second measurement possible calibration errors cannot be ascribed to uncertainties in the transmittance of the Baader OD5.0 solar filter employed for the first measurement. Because the two above absolute calibration methods are independent, and because the resulting disagreement with respect to absolute $pB$ values provided by MLSO is about the same, this suggests the possible existence of a systematic error in the measurements derived here from TSE images, whose origin is unknown.

\section{Derivation of electron densities} \label{sec: densities}

Once the $pB$ images are calibrated in standard units, the resulting values can be fitted radially to derive radial profiles of the coronal electron density $n_e$. This has been done here with the standard Van de Hulst inversion technique \citep{vandehulst1950}. This well-established method, assuming a simple spherically symmetric corona at each latitude, has been more recently validated by making a comparison with tomographic electron density reconstructions \citep{wang2014}.

The resulting 2D electron density maps as obtained with pB images for the three RGB channels are shown in \ref{Fig15}. These maps were obtained starting from $pB$ measurements derived with TSE images and after relative radiometric calibration to re-scale the observed values to those provided by MLSO in standard units of MSB. Because the absolute radiometric calibrations derived here with full Sun image and with stars give values of MSB a factor $\sim 2$ larger, these higher MSB values would simply reduce the $pB$, and thus the resulting densities, exactly by the same factor.
\begin{figure}[h!]
\centering
\includegraphics[width=0.48\textwidth]{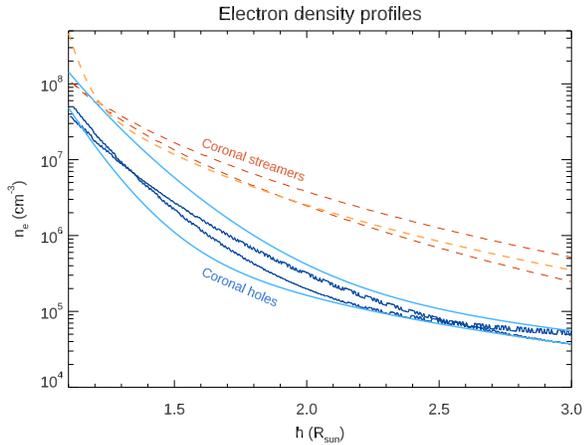}
\caption{
Electron density radial profiles plotted along coronal streamers (dashed red lines) and coronal holes (solid blue lines); the measured profiles are compared here with standard reference profiles given by \citet{gibson1999} for minimum coronal streamers (thicker orange dashed line) and by \citet{cranmer1999} and \citet{guhathakurta1999} for coronal holes (thicker cyan solid lines).
}
\label{Fig16}
\end{figure}

A more quantitative comparison is provided in Fig.~\ref{Fig16}, showing the electron density values derived here between 1.1 and 3 R$_{sun}$ in the two brighter coronal streamers (dashed red lines) visible in the top right and bottom left quadrants in Fig.~\ref{Fig15}, and in the nearby polar coronal holes (solid blue lines). Fig.~\ref{Fig16} also shows that the density profiles form TSE images have a very good agreement with reference values provided for instance by \citet{gibson1999} for minimum coronal streamers (thicker orange dashed line), and with values provided by \citet{cranmer1999} and \citet{guhathakurta1999} for coronal holes (thicker cyan solid lines).

\section{Summary and conclusions} \label{sec: summary}

In this work I analyzed the sequence of images acquired during the total solar eclipse (TSE) of August 21st, 2017 from the Idaho Falls area. The images were acquired with a standard DSLR camera, mounted on a simple fixed tripod, and equipped with a cost effective zoom and linear polarizing filter. After demosaicing to separate pixels in the three RGB colors of the Bayer filter matrix, the images (having a projected pixel size of 3.7 arcsec) were corrected for the dark currents and flat-field, and then co-aligned based on the detected position of the brighter star $\alpha-$Leonis. From bi-Gaussian fitting of the star intensity distribution, it turns out that the images have an effective resolution of about 10 acsecs, comparable with the apparent sky motion in 1 sec.

After image co-alignement, each sequence of bracketing images has been combined, by measuring the intensity for each pixel in the interval of linearity of the detector response as a function of the exposure time. This provided one mosaic image for each one of the three RGB colors: one triplet without a polarizer, and the second and third ones with a linear polarizer. Comparisons among radial intensity profiles obtained with pixels corresponding to different colors, hence located in different positions in the RGB Bayer filter matrix of DSLR camera, show considerable (up to $\sim 5-10$ \%) relative intensity differences in the inner coronal regions (below $\sim 1.5$ R$_{sun}$). This means that even small differences in the projected altitudes of nearby pixels in the camera (3.7 arcsecs, corresponding to $3.9 \times 10^{-3}$ R$_{sun}$) results in considerable relative intensity differences of the observed corona. This may partially affect the measurements of degree of linear polarization as obtained from a combination of intensities observed in nearby pixels with different orientations of linear polarizers by assuming that the different pixels are sampling the same corona, as done in the analysis of images acquired with a PolarCam \citep[e.g.][]{reginald2017, judge2019, fineschi2019}. In principle, images acquired with PolarCams should be analyzed instead with methods similar to those developed by many authors for debayering or demosaicing regular RGB images acquired by DSLR cameras \citep[e.g.][]{rajeev2002, parmar2005}, in particular if one wants to exploit these images to resolve fine features located in the inner corona.

In this work, the mosaic images acquired with different polarizations have been combined, to measure the degree of linear polarization $p$ [\%], and the three components [I,Q,U] of the Stokes vector. Despite the presence of a few artifacts (ghosts due to reflections from the polarizing filter, and a cross-like divergent pattern where solutions for the Q-component of Stokes vector diverge), the resulting values of $p$ are in nice agreement with those provided for instance by \citet{snik2020} for the same TSE. Relative radiometric calibration has been performed re-scaling measurements of polarized brightness $pB$ obtained here from from TSE with those provided by the Mauna Loa K-Cor coronagraph (MLSO), showing a very good agreement both in the latitudinal distribution and at different altitudes up to 2 R$_{sun}$, with very good signal-to-noise ratio.

Absolute radiometric calibrations of $pB$ images have been performed as well, with two different methods: by measuring the full disk Mean Solar Brightness (MSB) with one image acquired before the partial eclipse with a OD5.0 filter, and by measuring the brightness of $\alpha-$Leo and $\nu-$Leo stars visible in the eclipse images. Both methods provided values of MSB approximately a factor $\sim 2$ larger that what derived from inter-calibration with MLSO $pB$ measurements, resulting in coronal densities lower by a factor $\sim 2$ than what could be derived from MLSO calibrated images. The reason for this systematic disagreement is unknown. It is curious to notice here that a similar systematic disagreement by about a factor $\sim 2$ was recently found also by \citet{lamy2019}, from a comparison between the $pB$ values measured by MLSO and LASCO on the one hand, and those predicted from MHD numerical simulations on the other hand (see their Fig. 23, left panels), with MLSO $pB$ values higher again than those predicted by MHD simulation, something that these authors ascribed to the possible uncertainties in coronal abundances, affecting the radiative loss function. Finally, $pB$ measurements have been employed here to derive an electron density image, and resulting values are in agreement with those measured by previous authors in coronal streamers \citep{gibson1999} and coronal holes \citep{cranmer1999, guhathakurta1999}. This is the first published map of coronal electron density measurements for the August 21st, 2017 TSE. 

In summary, this work demonstrates that images acquired during TSE with cost-effective amateur equipment can provide high-quality images that can be employed for scientific analysis purposes. In the aim of the author, this work will hopefully inspire and motivate future amateur astronomers and educators to create projects based on similar images acquired with DSLR cameras during TSEs, for instance in the occasion of the forthcoming eclipse on 8 April 2024 that will cross again the US and will last more than 4 minutes.

\acknowledgments
The author thanks L. Abbo and C. Benna for their invaluable help and support in the organization of the observational campaign, making this work possible. The author also thanks the anonymous Referee for useful suggestions.

\bibliographystyle{aasjournal}
\bibliography{biblio2}{}

\end{document}